\documentstyle[12pt]{article}
\def\Re{{\cal R \mskip-4mu \lower.1ex \hbox{\it e}\,}}
\def\Im{{\cal I \mskip-5mu \lower.1ex \hbox{\it m}\,}}
\def\snu{\ifmmode \tilde\nu \else $\tilde\nu$\fi}
\def\slep{\ifmmode \tilde l \else $\tilde l$\fi}
\def\ie{{\it i.e.}}
\def\eg{{\it e.g.}}
\def\etc{{\it etc}}
\def\etal{{\it et al.}}

\def\sub#1{_{\lower.25ex\hbox{$\scriptstyle#1$}}}
\def\sul#1{_{\kern-.1em#1}}
\def\sll#1{_{\kern-.2em#1}}
\def\sbl#1{_{\kern-.1em\lower.25ex\hbox{$\scriptstyle#1$}}}
\def\ssb#1{_{\lower.25ex\hbox{$\scriptscriptstyle#1$}}}
\def\sbb#1{_{\lower.4ex\hbox{$\scriptstyle#1$}}}

\def\to{\rightarrow}
\def\mh{\ifmmode m\sbl H \else $m\sbl H$\fi}
\def\mch{\ifmmode m_{H^\pm} \else $m_{H^\pm}$\fi}
\def\mt{\ifmmode m_t\else $m_t$\fi}
\def\mc{\ifmmode m_c\else $m_c$\fi}
\def\mz{\ifmmode M_Z\else $M_Z$\fi}
\def\mw{\ifmmode M_W\else $M_W$\fi}
\def\mws{\ifmmode M_W^2 \else $M_W^2$\fi}
\def\mhs{\ifmmode m_H^2 \else $m_H^2$\fi}
\def\mzs{\ifmmode M_Z^2 \else $M_Z^2$\fi}
\def\mts{\ifmmode m_t^2 \else $m_t^2$\fi}
\def\mcs{\ifmmode m_c^2 \else $m_c^2$\fi}
\def\mchs{\ifmmode m_{H^\pm}^2 \else $m_{H^\pm}^2$\fi}
\def\ztwo{\ifmmode Z_2\else $Z_2$\fi}
\def\zone{\ifmmode Z_1\else $Z_1$\fi}
\def\mtwo{\ifmmode M_2\else $M_2$\fi}
\def\mone{\ifmmode M_1\else $M_1$\fi}
\def\tb{\ifmmode \tan\beta \else $\tan\beta$\fi}
\def\xw{\ifmmode x\sub w\else $x\sub w$\fi}
\def\ch{\ifmmode H^\pm \else $H^\pm$\fi}
\def\lum{\ifmmode {\cal L}\else ${\cal L}$\fi}
\def\inpb{\ifmmode {\rm pb}^{-1}\else ${\rm pb}^{-1}$\fi}
\def\infb{\ifmmode {\rm fb}^{-1}\else ${\rm fb}^{-1}$\fi}
\def\epem{\ifmmode e^+e^-\else $e^+e^-$\fi}
\def\ppb{\ifmmode \bar pp\else $\bar pp$\fi}

\newskip\zatskip \zatskip=0pt plus0pt minus0pt
\def\matth{\mathsurround=0pt}

\def\atversim#1#2{\lower0.7ex\vbox{\baselineskip\zatskip\lineskip\zatskip
  \lineskiplimit 0pt\ialign{$\matth#1\hfil##\hfil$\crcr#2\crcr\sim\crcr}}}

\renewcommand{\thefootnote}{\fnsymbol{footnote}}

\begin{document} \begin{titlepage}
\setcounter{page}{1}
\thispagestyle{empty}
\rightline{\vbox{\halign{&#\hfil\cr
&ANL-HEP-PR-92-81\cr
&September 1992\cr}}}
\vspace{1in}
\begin{center}

{\Large\bf
Lepton Number Violating Radiative $W$ Decay in Models with R-parity
Violation}
\footnote{Research supported by the
U.S. Department of
Energy, Division of High Energy Physics, Contract W-31-109-ENG-38.}
\medskip

\normalsize THOMAS G. RIZZO
\\ \smallskip
High Energy Physics Division\\Argonne National
Laboratory\\Argonne, IL 60439\\

\end{center}

\begin{abstract}

Models with explicit R-parity violation can induce new rare radiative decay
modes of the $W$ boson into single supersymmetric particles which also violate
lepton number.
We examine the rate and signature for one such decay, $W\rightarrow
\tilde l\gamma$, and find that such a mode will be very difficult to observe,
due its small branching fraction,
even if the lepton number violating coupling in the superpotential is
comparable in strength to electromagnetism. This parallels a similar result
obtained earlier by Hewett in the case of radiative $Z$ decays.

\end{abstract}

\renewcommand{\thefootnote}{\arabic{footnote}} \end{titlepage}

%%%%%%%%%%%%%%%%%%%%%%%%%%%%%%%---- text

In the simple Minimal version of the Supersymmetric~Standard~Model(MSSM), both
baryon($B$) and lepton($L$) numbers are simultaneously conserved quantities
due to the imposition of a discrete symmetry called R-parity. One can then
assign to each Standard Model(SM) particle and its SUSY partner a
multiplicatively conserved
quantum number, given by $R=(-1)^{2S+3B+L}$, where $S$ is just the particle's
spin. In addition to
the gauge symmetries and the assumption of minimal particle content, R-parity
conservation severly limits the possible interactions among the usual SM
fermions and their superpartners as well as the properties of the SUSY
partners themselves. The two most important phenomenological consequences
{\cite {susyrefs}} are
well-known: ({\it {i}}) The SUSY partners of the conventional SM particles
which carry negative R-parity can only be produced
in pairs and ({\it {ii}}) the lightest supersymmetric particle(LSP), is an
electrically neutral,
color-singlet and is stable. This second property is the one which is
conventionally used
experimentally to search for SUSY particles, \ie, once they are pair-produced
their decays involve final
states which include the LSP that only appears as a missing energy signature
in a collider detector. If R-parity were broken both these conclusions would
be invalidated leading to an entirely new phenomenology.

Of course this minimal approach may not be that realized by nature. In
particular, it is possible to construct phenemenologically viable models
wherein R-parity is violated either spontaneously{\cite {spon}}, through the
acquisition of a vacuum expectation value(vev) by a sneutrino, or explicitly
via the existence of additional terms in the superpotential{\cite {expl}}
constructed out of the conventional superfields. These
additional terms are possible as the gauge symmetries {\it alone} do not
forbid their existence. If such new interactions are present they can lead not
only to a destablization of the LSP but also new production modes for SUSY
partners not present in the MSSM and thus forces us to a re-evaluate of the
traditional
search techniques for these particles. Clearly, the breaking of R-parity
implies that these
new interactions will violate $L$ and $B$. In such a R-parity breaking
scenario this
more general form of the superpotential, $W$, can be written as
\begin{eqnarray}
W= & h_{ij}L^i_L H_2\bar E^j_R+h_{ij}'Q^i_L H_2\bar D^j_R
+h_{ij}''Q^i_L H_1\bar U^j_R \nonumber \\
& +\lambda_{ijk}L^i_L L^j_L \bar E^k_R +\lambda_{ijk}'L^i_L Q^j_L\bar D^k_R \\
& +\lambda_{ijk}'' \bar U^i_R\bar D^j_R\bar D^k_R \,, \nonumber
\end{eqnarray}
where $ijk$ are generation indices, and the $h$'s and $\lambda$'s are
{\it {a priori}} unknown Yukawa couplings. $Q_L,L_L$, \etc ~represent the usual
left-handed chiral superfields of the MSSM. Whereas the $h$ couplings are the
standard ones responsible for generating fermion masses when the scalar
components of $H_1$ and $H_2$ acquire vev's, the $\lambda$, $\lambda'$, and
$\lambda''$ terms lead to generation-dependent $\Delta L$ and $\Delta B$
interactions. Of course, if all such terms are allowed {\it {simultaneously}},
proton decay proceeds unsuppressed and thus, both $\Delta L$ and $\Delta B$
terms cannot be present. For example, if $\lambda$ = $\lambda'$ =0, only
$B$ violation would occur. In particular, it has be recently shown
{\cite {bento}} that if one
assumes only the particle content of the MSSM and the lack of rapid proton
decay, then two unique discrete symmetries are possible: $R$ and $B$, the
so-called baryon-parity. Thus the most
reasonable scenario to consider for our purposes is one in which $B$ is
conserved while both $R$ and $L$ are violated; this corresponds to setting
all of the $\lambda''$ terms to zero in $W$. In principle, this class of
models should be considered as to be just as likely a scenario for the
realization of SUSY as is the more conventional MSSM with the same particle
content. This is the scheme we consider below.

The interaction Lagrangians that result from $W$ can be written as
\begin{equation}
{\cal L} =  \lambda_{ijk}[\snu^i_L\bar e^k_Re^j_L+\tilde e^j_L\bar e^k_R\nu^i_L
+(\tilde e^k_R)^*(\bar\nu^i_L)^ce^j_L  -(i\leftrightarrow j)] + h.c.
\end{equation}
and
\begin{eqnarray}
{\cal L} = & \lambda_{ijk}'[\snu^i_L\bar d^k_Rd^j_L+\tilde d^j_L\bar d^k_R
\nu^i_L+(\tilde d^k_R)^*(\bar\nu^i_L)^cd^j_L \\
& -\tilde e^i_L\bar d^k_Ru^j_L-\tilde u^j_L\bar d^k_Re^i_L
-(\tilde d^k_R)^*(\bar e^i_L)^cu^j_L] + h.c. \nonumber
\end{eqnarray}
Once B-parity has been imposed, many of the remaining couplings can be
constrained via
various phenomenological considerations using these Lagrangians as has been
done by several groups
of authors{\cite {bigref}}. Combining all of their results, one finds that the
only potentially large couplings, \ie, ones that are capable of being of the
same strength as electromagnetism, are{\cite {godb}}: $\lambda_{131}$,
$\lambda'_{3jk}$, $\lambda'_{121}$, $\lambda'_{222}$, $\lambda'_{223}$,
$\lambda'_{232}$, and $\lambda'_{233}$.  Note particularly that interactions
involving third generation particles are least likely to be constrained by
existing data.

Although one can look for the direct influence of R-parity violating
interactions, it may be possible to search for their indirect effects through,
\eg, loop diagrams. One such possiblility, recently examined by Hewett
{\cite {jlh}}, is the loop-induced decay $Z\to \snu\gamma$, with a rate that
might be only an order of magnitude smaller than that for $Z\to H\gamma$ in
the SM.  In this paper we wish to examine the corresponding process in
$W^{\pm}$ decay, \ie, $W\to \slep\gamma$. Unfortunately, as we will see, the
branching fraction for this process is found to be much smaller than in the
$Z$ case due to helicity supression and the form of the superpotential, $W$.

The diagrams responsible for the $W\to \slep\gamma$ decay are shown in Fig.~1
and involve one R-parity violating vertex. For simplicity, we have assumed
that only one of the two $\Delta L$ couplings is non-zero, \ie, we ignore the
possible contributions of the $\lambda$ terms in the superpotential and
concentrate on the $\lambda'$ terms since more of them can be large. We note,
however, that since $\lambda'_{131}$ can be sizeable, a loop involving first
generation leptons might yield a significant contribution. In fact, one finds
that such contributions are suppressed due to the small masses of these
particles. This being the case, it is the couplings $\lambda'_{233}$ and
$\lambda'_{333}$ which are relevant here. In the former case, the $\slep$ is a
smuon whereas in the latter it is a stau. We, of course, do not know the size
of this active $\lambda'$ coupling so we simply scale it to the electomagnetic
strength as is customary, \ie, we write the effective $tb\slep$ interaction as
\begin{eqnarray}
{\cal L} = & -e{\sqrt {F}}~\tilde l_L\bar b(1-\gamma_5)t
\end{eqnarray}
where $F$ is an unknown parameter which may be as large as unity. With this
normalization, the amplitude for the $W\to \slep\gamma$ decay process can be
written as
\begin{equation}
{\cal A}=\Big[F_1{(q_\nu k_\mu-g_{\nu\mu}k\cdot q)\over M_W^2} +
iF_2{\epsilon_{\mu\nu\sigma\tau}q^\sigma k^\tau\over M_W^2}\Big]
\epsilon^\mu_\gamma \epsilon^\nu_W
\end{equation}
with q(k) being the momentum of the photon($W$). In terms of the form factors
$F_{1,2}$, the decay width is given by
\begin{equation}
\Gamma(W\to \slep\gamma)={M_W^3 \over {96\pi^2}}(F_1^2+F_2^2)
\Big(1-{m_{\slep}^2 \over {M_W^2}}\Big)^3
\end{equation}
with $m_{\slep}$ being the slepton mass. Definining the mass difference,
$\delta=m_{\slep}^2-M_W^2$, we find that $F_{1,2}$ can be written as
\begin{eqnarray}
F_1 &={-i\alpha gN_c m_b \sqrt {F} \over \sqrt {2}\pi\delta}
[Q_uI_1+Q_dI_2] \\
F_2 &={-i\alpha gN_c m_b \sqrt {F} \over \sqrt {2}\pi\delta}
[Q_uI_3+Q_dI_4] \nonumber
\end{eqnarray}
where $N_c$=3 is the usual color factor, $m_b$ is the b-quark mass, $g$ is the
conventional weak coupling constant, $Q_{u,d}$ are the electric charges of
the up- and down-quarks, and $I_i$ can be expressed as sums of
parameter integrals:
\begin{eqnarray}
I_1 & = & 1+2m_t^2 \delta^{-1} G_{-1}(m_t,m_b)
+2[\delta^{-1}(m_b^2-m_t^2-M_W^2)
-1/2] G_0(m_t,m_b) \nonumber \\
    & + & 2M_W^2 \delta^{-1} G_1(m_t,m_b) \nonumber \\
I_2 & = & 1+2m_b^2 \delta^{-1} G_{-1}(m_b,m_t)
+2[\delta^{-1}(m_t^2-m_b^2-M_W^2)
-1/2] G_0(m_b,m_t) \nonumber \\
    & + & 2M_W^2 \delta^{-1} G_1(m_b,m_t) \\
I_3 & = & -G_0(m_t,m_b) \nonumber \\
I_4 & = & G_{-1}(m_b,m_t)-G_0(m_b,m_t) \nonumber
\end{eqnarray}
where the $G_n$ are given by
\begin{equation}
G_n(m_i,m_j)=\int^1_0 dz z^n \ln\Bigg[{m_i^2(1-z)+m_j^2z-z(1-z)m_{\slep}^2
\over
m_i^2(1-z)+m_j^2z-z(1-z)M_W^2} \Bigg]
\end{equation}
It is important to note that both $F_{1,2}$ are proportional to $m_b$ and not
$m_t$. This comes about due to the fact that the $W$ charged current
interactions are purely left-handed, as is the coupling in Eq.(4), and that
unless a mass term from a propagator of an internal quark line is picked up,
the result will vanish since a trace over an odd number of $\gamma$-matrices
will then be taken. In the actual calculation only the b-quark mass term
is picked
up so that instead of increasing in magnitude as $m_t$ increases, the rate for
$W \to \slep \gamma$ will decrease for large $m_t$ due to supression from the
top-quark propagator.

Using $m_b$=5 GeV, $\alpha^{-1}$=127.9, and $M_W$=80.15 GeV as numerical
input, we can calculate the partial decay width as a function of $m_{\slep}$
for different values of $m_t$ and the parameter $F$. The result of this
calculation, expressed as the branching fraction(B) for the
$W \to \slep \gamma$ process, is shown in Fig.2. As advertised, even for
$F$=1, we see that B is less than about $10^{-8}$ for all values of $m_t$ and
$m_{\slep}$ and, as anticipated, falls with increasing $m_t$. This rate is
sufficiently tiny that this reaction will be impossible to observe. If the
helicity structure of the R-parity violating interaction had been opposite,
the rate could have
been larger by a factor of $m_t^2/m_b^2 \simeq 10^3$ and potentially
observable since the overall nuerical factor in the amplitude would then have
been proportional to $m_t$ instead of $m_b$.

Unfortunately, our result implies that signals for R-parity violating
interactions must be sought elsewhere than in radiative $W$ decays.

\newpage
\vskip.25in
\centerline{ACKNOWLEDGEMENTS}

The author would like to thank J.L. Hewett for discussions related to this
work.This research was supported in part by the U.S.~Department of Energy
under contract W-31-109-ENG-38.

\newpage

%
%%%%%%%%%%%%%%%%%%--- References
%%%%%%%%%%%%%%%%%%%%%%%%%%%%%%%%%%%%%%%%%%%%%%%%%%%%%%%
\def\MPL #1 #2 #3 {Mod.~Phys.~Lett.~{\bf#1},\ #2 (#3)}
\def\NPB #1 #2 #3 {Nucl.~Phys.~{\bf#1},\ #2 (#3)}
\def\PLB #1 #2 #3 {Phys.~Lett.~{\bf#1},\ #2 (#3)}
\def\PR #1 #2 #3 {Phys.~Rep.~{\bf#1},\ #2 (#3)}
\def\PRD #1 #2 #3 {Phys.~Rev.~{\bf#1},\ #2 (#3)}
\def\PRL #1 #2 #3 {Phys.~Rev.~Lett.~{\bf#1},\ #2 (#3)}
\def\RMP #1 #2 #3 {Rev.~Mod.~Phys.~{\bf#1},\ #2 (#3)}
\def\ZP #1 #2 #3 {Z.~Phys.~{\bf#1},\ #2 (#3)}
\def\IJMP #1 #2 #3 {Int.~J.~Mod.~Phys.~{\bf#1},\ #2 (#3)}

\newpage

%%%%%%%%%%%%%%%%%%%%%%%--- figures
%
{\bf Figure Captions}
\begin{itemize}

\item[Figure 1.]{The Feynman diagrams responsible for $W\to\slep\gamma$
in R-parity violating models.}
\item[Figure 2.]{The branching fraction, B, for the decay $W\to\slep\gamma$ as
a function of the slepton mass, $m_{\slep}$, assuming $F$=1 for various values
of the top
quark mass: $m_t$=100 (150, 200) GeV corresponds to the solid (dash-dotted,
dashed) curve.}
\end{itemize}


\begin{thebibliography}{99}
\bibitem{susyrefs}
For an introduction and phenomenological overview of SUSY, see H.E.\ Haber and
G.L.\ Kane, \PR 117 75 1985 .
\bibitem{spon}
M.C.\ Gonzalez-Garcia, J.C.\ Romano, and J.W.F.\ Valle, Univeristy of
Valencia report FTUV-91-42 1991); J.C.\ Romao, C.A.\ Santos, and
J.W.F.\ Valle, \PLB B288 311 1992 ; J.C.\ Romao and J.W.F.\ Valle,
\NPB B381 87 1992 ; A.\ Masiero and J.W.F.\ Valle, \PLB B251 273 1990 ;
A.\ Santamaria and J.W.F.\ Valle, \PRD D39 1780 1989 and \PLB B195 423 1987 ;
J.C.\ Romao, F.\ de\ Campos, and J.W.F.\ Valle, \PLB B292 329 1992 .
\bibitem{expl}
L.J.\ Hall and M.\ Suzuki, \NPB B231 419 1984 ;
S.\ Dawson, \NPB B261 297 1985 ;
S.\ Dimopolous and L.S.\ Hall, \PLB B207 210 1987 ;
H.\ Dreiner and G.\ Ross, \NPB B365 597 1991 and Oxford University report
OUTP-92-08P (1992);
R.\ Mohapatra, \PRD D34 3457 1989 ;
E.\ Ma and D.\ Ng, \PRD D41 1005 1990 ;
M.\ Doncheski and J.L.\ Hewett, Argonne National Laboratory report ANL-HEP-PR-
92-28 (1992), to appear in Z.\ Phys.\ {\bf C}.
\bibitem{bento}
M.C.\ Bento, L.J.\ Hall, and G.G.\ Ross, \NPB B292 400 1987 ; see also the
paper by Dreiner and Ross in Ref.2.
\bibitem{bigref}
See, for example, V. Barger, G.F. Giudice, and T. Han, \PRD D40 2987 1989 ;
C.S.\ Aulakh and R.N.\  Mohapatra, \PLB B119 316 1982 ;
L.J.\ Hall, \MPL A5 467 1990 ;
S.\ Dimopoulos \etal , \PRD D41 2099 1990 ;
R.\ Arnowitt and P.\ Nath in {\it Phenomenology of the Standard Model and
Beyond},eds by D.P.\ Roy and P.\ Roy, World Scientific, Singapore, 1989.
\bibitem{godb}
R.M.\ Godbole, P.\ Roy, and X.\ Tata, CERN report CERN-TH.6613/92 (1992).
\bibitem{jlh}
J.L.\ Hewett, Argonne National Laboratory report ANL-HEP-CP-92-23 (1992),
and talk given at the {\it Workshop on Photon Radiation From Quarks}, Annecy,
France, December 2-3, 1991.

\end{thebibliography}
\end{document}